# Frenkel anomaly on co-ordination numbers in liquid $CO_2$ at 100 and 1000 bar studied by Monte Carlo simulation using Kihara potential model

Koji Kobashi

Former Research Assistant, Physics Department, Colorado State University, Fort Collins, CO, USA,
and Former Senior Researcher, Kobe Steel, Ltd., Japan

**Abstract**

An issue concerning the Frenkel line of liquid $CO_2$ is that its location has not been unequivocally determined. So far, reliable Frenkel lines were identified from velocity autocorrelation functions (VAFs) computed by Molecular Dynamic simulations; however VAFs cannot be directly verified experimentally. By contrast, the co-ordination numbers ($CNs$) can be measured experimentally by X-ray and neutron scattering or computed by Monte Carlo (MC) simulation, and thus provide an alternative means of determining the location of the Frenkel line. In the present study, the $CNs$ were computed by MC simulation using the Kihara potential, and the Frenkel anomalies were identified at $\{P = 100$ bar, $T = 260$ K$\}$, in good agreement with previous results based on VAFs, and at $\{P = 1000$ bar, $T = 370$ K$\}$, which deviates significantly from them.

**I. Introduction**

On a pressure-temperature (*P-T*) phase diagram of small molecules such as Ne, Ar, $N_2$, $H_2O$, $CH_4$, and $CO_2$, the liquid region adjacent to the melting curve exhibits temporal characteristics similar to solid, and is called a *rigid liquid*. The rigid liquid transforms either into a normal or into a supercritical (SC) liquid (both of which will hereafter be referred to simply as a *normal fluid*) across a narrow region away from the melting curve. This phenomenon was first discovered by Frenkel [1], and is referred to in the present article as the *Frenkel anomaly*. The narrow boundary separating the rigid liquid from the normal fluid is known as the Frenkel line [2]. The existence of the Frenkel line has been experimentally and theoretically confirmed for Ne, Ar, $N_2$, $H_2O$, $CH_4$, and $CO_2$. as well as Lennard-Jones (LJ) fluids [2-10]. On the phase diagram, the Frenkel line starts from {*T* = 0.7 to 0.8 $T_c$ and *P* = 0.7 to 0.8 $P_c$} [10, 11], where $T_c$ and $P_c$ are the critical *T* and *P*, respectively, of the liquid under consideration, and extends into the SC liquid region at high *T* and *P*. According to Frenkel [1], each molecule in a liquid is oscillating around its equilibrium position and jumps to a neighboring position in time *τ*. Thus, the time *τ* represents the average time between two consecutive events of molecular diffusion (or structural relaxation). This is compared with the minimal oscillation period $\tau_D$ (≈ 0.1 to 1 ps: this period is called the (minimal) Debye oscillation period) in the rigid liquid [2, 10]. Brazhkin *et al*. summarized [2] that (i) $\tau > \tau_D$ in the rigid liquid, (ii) $\tau < \tau_D$ in the normal fluid, and (iii) $\tau \approx \tau_D$ on the Frenkel line, meaning that the Frenkel anomaly is dynamical in nature, and crossing the Frenkel line leads to changes in physical properties such as sound (velocity and shear waves), diffusion, viscosity, and thermal conductivity [3]. The Frenkel anomaly is often identified through the velocity autocorrelation function (VAF) [2, 3, 10, 11]. In the rigid liquid, the VAF shows a pronounced minimum due to molecular oscillations, which disappears when the Frenkel line is crossed into the normal fluid. Such a change has been observed in molecular dynamics (MD) simulations [7-11].

The Frenkel anomaly is associated not only with different dynamical behaviors of liquids between the rigid liquid and the normal fluid [2, 4] but also with static properties. A well-known property is the heat capacity at constant volume, $c_V$, along the Frenkel line. For monoatomic liquids, the heat capacity $c_V = 3\ k_B$ in the rigid fluid, while $c_V = 2k_B$ immediately after the Frenkel line is crossed ($k_B$ is the Boltzmann constant). This simple criterion holds surprisingly well for He, Ne, and Ar. Raman [3, 5, 8] and infrared spectroscopy is also known to exhibit noticeable changes in the integrated intensities and the widths of intramolecular vibrational bands when the Frenkel line is crossed [8]. Another static property, which is the main subject of the present study, is the co-ordination number (*CN*) [3, 4, 8, 12] that is the average number of molecules in the first shell surrounding a central molecule. In solid $CO_2$ phase I, *CN* = 12; in rigid liquid, it is around 12 (11 to 13); and in a normal fluid and gas, it is smaller. The *CN* drops quickly in crossing the Frenkel line from a rigid liquid to a normal fluid [3], and this phenomenon is experimentally confirmed by X-ray and small-angle neutron diffractions (SANS) [3, 4, 8]. The definition of *CN* will be described in more detail later in Sec. 4. An issue of liquid $CO_2$ is that the location of the Frenkel line has not been unequivocally determined despite many excellent works [3, 13-16] and there exist significant disagreement between theory (computer simulations) and experiment and even among the results of simulations. Most importantly, the cause of the Frenkel anomaly has not yet been understood even though it is really a peculiar phenomenon.

In the precedent study by the present author [17], the radial distribution function (RDF) of liquid $CO_2$ at $\{P = 50$ bar, $T = 250$ K$\}$ and its expansion coefficients by spherical harmonics were computed for all possible combinations of angular indices $l$, $l'$, and $|m| \leq 8$ by Monte Carlo (MC) simulation using the Kihara potential model. The present simulation was based on this work as well as excellent text books [3, 18-21]. In the present study, the locations of the Frenkel anomalies were determined from *CNs* computed sby MC simulation. As shown on the phase diagram of $CO_2$ (Fig. 1), the $\{P, T\}$ ranges were selected so that it crosses the Frenkel line determined in Refs. 2, 3, and 10. The $\{P, T\}$ ranges selected were $\{P = 100$ bar, $T = 220 - 300$ K: the normal liquid region$\}$ and $\{P = 1000$ bar, $T = 240 - 420$ K: the SC liquid region$\}$. In Fig. 1, the Frenkel line reported in Ref. 10 is depicted by a dark-blue dashed line. The green dashed line is the Frenkel line experimentally determined by Pipich and Schwahn [22, 23]. The $\{P, T\}$ conditions used for the MC simulations in the present study are shown by the red closed circles, and the Frenkel anomalies obtained are depicted by the red open circles.

Next Sec. 2 describes the Kihara potential model, and Sec. 3 presents the computational procedure of the MC simulation employed in the present study. These items are described as briefly as possible because the detailed descriptions have been given in Ref. 17. Section 4 presents results and discussion on *CNs* for $P = 100$ bar and 1000 bar. Section 5 describes the molecular orientation in liquid $CO_2$. The conclusion is given in the final Sec. 6.

So far, the most reliable Frenkel line seems to be the one determined by VAF simulation in Ref. 10, and it is quoted in a textbook [3]. The main purpose of the present study is to compute the *CNs* by MC simulation to identify Frenkel anomalies and compare the results with those reported in Ref. 10. As a result, Frenkel anomalies were identified at $\{P = 100$ bar, $T = 260$ K$\}$ that was in good agreement with the results of the Frenkel line [10] and at $\{P = 1000$ bar, $T = 370$ K$\}$ that was significantly in disagreement with the result reported in Ref. 10. In conducting the present research of liquid $CO_2$, useful works using X-ray and neutron scattering as well as computer simulations [3, 13-15] have been consulted.

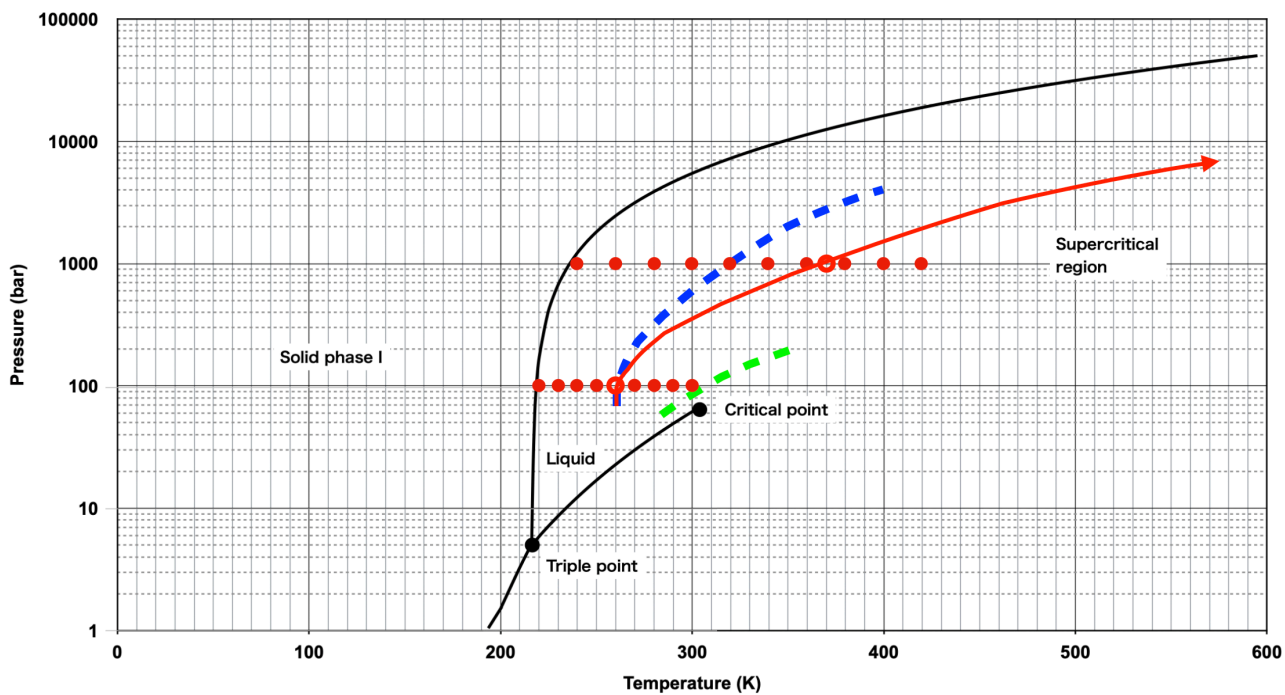

Fig. 1. Phase diagram of $CO_2$. The nine red points at $P$ = 100 bar indicate the $\{P, T\}$ conditions used in the present MC simulations, and the red open circle is where the Frenkel anomaly was identified in the present study. On the other hand, the ten red closed points at $P$ = 1000 bar indicate the $\{P, T\}$ conditions used in the present MC simulations, and the red circle is where the Frenkel anomaly was identified. The red arrow is a Frenkel line that linked the two red open points smoothly and extended to high $P$ and $T$ on the phase diagram. The dark-blue line is the Frenkel line presented in Refs. 3 and 10, while the green dashed line is the Frenkel line determined in Ref. 22. Note that the triple point of $CO_2$ is $\{P$ = 5.2 bar, $T$ = 216.6 K$\}$ and the critical point is $\{P_c$ = 73.8 bar, $T_c$ = 304.1 K$\}$. See Ref. 24 for the phase boundaries denoted with dark lines, and Ref. 25 for $\{P, T\}$ and other parameters.

**2. Kihara core potential and Kihara potential**

The Kihara core potential assumes a solid core inside a molecule, and defines the intermolecular potential as a function of the nearest core-core distance, $\rho$, as shown in Fig. 2. The core is a rod with a length of $l$ = 2.21 Å (10 Å = 1 nm) and zero diameter.

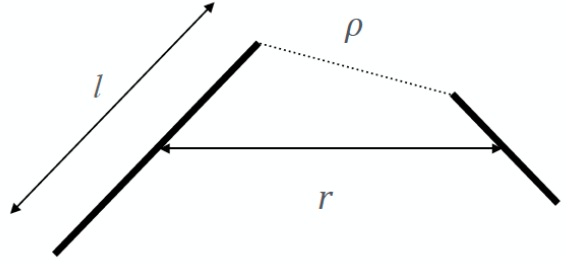

Fig. 2. Concept of Kihara core potential for $CO_2$. $l$: core length, $r$: molecular center-to-center distance, and $\rho$: the shortest core-core distance [17, 26]
.

The Kihara potential between $CO_2$ molecules consists of a core-core potential and a point electrostatic quadrupole-quadrupole (EQQ) potential. The Kihara core-core potential is expressed by a LJ potential $V_c(\rho)$:

$$V_c(\rho) = U_0 \left[ \frac{6}{n-6} \left(\frac{\rho_0}{\rho}\right)^n - \frac{n}{n-6} \left(\frac{\rho_0}{\rho}\right)^6 \right]. \qquad (1)$$

In the present study, $n = 9$, $U_0 = 250$ K (1 K = 1.38 x 10$^{-23}$ J), $\rho_0 = 3.27$ Å, and $l = 2.21$ Å. Note that the core length $l$ was smaller than the experimental O-O distance, 2.30 Å. In the actual computations, the maximum distance of the Kihara core-core potential was limited to $r = 10$ Å, and only the van der Waals potential of Eq. 2 that has no orientation dependence was applied between molecular centers beyond $r = 10$ Å up to $r = 15$ Å:

$$V_c(\rho) = -U_0 \left[ \frac{n}{n-6} \left(\frac{\rho_0}{r}\right)^6 \right] . \qquad (2)$$

The EQQ potential between $CO_2$ molecules was expressed by a point electrostatic quadrupole moment, Q = - 4.8 x 10$^{-26}$ esu.cm$^2$, placed in the center of molecule and assumed to work up to $r = 15$ Å. For the *Pa3* structure of solid $CO_2$ phase I at ambient pressure and $T = 5$ K, where the lattice constant was $a = 5.54$ Å, the EQQ potential energy accounted for 43% of the total intermolecular potential energy, a similar magnitude of the sum of Kihara core potential energy and the van der Waals potential energy. The EQQ potential is strongly orientation dependent, and determines the molecular orientations in the *Pa3* structure.

### 3. Computational procedure of MC simulation [17]

A total of 2048 $CO_2$ molecules were contained in a cubic box of volume $V$ with the edge length $A_L$ ($A_L \approx 50$ Å, $V = A_L^3$), and the periodic boundary condition was applied in the potential energy computations. In all cases of the present MC simulation, $A_L \geq 15$ Å (the potential interaction distance). The molecular coordinates of the *i*-th molecule included the molecular center $r_i = \{x_i, y_i, z_i\}$ and the molecular orientation $\omega_i = \{\theta_i, \varphi_i\}$, where *i* runs from 1 through 2048. Following Ref. 18, $ln(V)$ was assigned to a random parameter for the volume. Thus, a MC trial was performed by randomly choosing either the *i*-th molecule or the volume $V$. If the *i*-th molecule was chosen, new coordinates of both molecular center $r_i'$ and the molecular orientation $\omega_i'$ were generated in such a way as $x_i' = x_i + (2q - 1) \Delta x_i$ and $\theta_i' = \theta_i + (2q - 1) \Delta \theta_i$, where $q$ is a random number, and both $\Delta x_i$, $\Delta y_i$ and $\Delta z_i$ ($\Delta x_i = \Delta y_i = \Delta z_i$ : these are denoted as $\Delta R_i = \{\Delta x_i, \Delta y_i, \Delta z_i\}$) and $\Delta \theta_i$ and $\Delta \varphi_i$ ($\Delta \theta_i \neq \Delta \varphi_i$ ; these are denoted as $\Delta \Omega_i = \{\Delta \theta_i, \Delta \varphi_i\}$) are external parameters. If the volume $V$ was chosen for the MC trial, a new volume $V'$ was generated according to $ln(V') = ln(V) + (2q - 1) \Delta V_m$, where $\Delta V_m$ is an external parameter. Thus, there were four external parameters in the *NPT*-MC simulation.

The total number of the MC trials was typically $1.5 \times 10^9$. For the external parameters, $\Delta R_i = 1.0$ Å initially and gradually decreased to 0.3 Å at the end. On the other hand, $\Delta \Omega_i = \{45°, 90°\}$ initially and gradually decreased to $\Delta \Omega_i = \{15°, 30°\}$ at the end. Similarly, $\Delta V_m = 0.2$ initially and gradually decreased to $\Delta V_m = 0.001$ at the end. In the final round of the *NPT*-MC simulation of $10^8$ MC trials, $A_L$, molar volume, and the total energy were stored each time a new volume was accepted, and their averages were taken after the computation was over. There were roughly 4500 data sets. The average values of $A_L$ and total energy were used for the following *NVT*-MC simulations along with last values of $\{r_i, \omega_i\}$. The average value of $A_L$ was slightly different from the last value of the *NPT*-MC simulation so that MC trials of $10^7$ were carried out using $\Delta R = 0.1$ Å and $\Delta \Omega = \{15°, 30°\}$ to relax the molecular configurations. Subsequently, 200 sets of *NVT*-MC simulations and a computation of the RDF was performed, in which each simulation consists of MC trials of $10^7$ with $\Delta R_i = 0.5$ Å and $\Delta \Omega_i = \{15°, 30°\}$. In the same program, 200 sets of RDF data were computed and averaged to obtain the final RDF.

### 4. Results and Discussion

The MC simulations were carried out for two pressures: the first was $P = 100$ bar and $T$ was from 220 to 300 K, in which $CO_2$ is the normal fluid. The second was $P = 1000$ bar and $T$ was from 240 to 420 K, in which $CO_2$ is the SC fluid. These states were selected from Ref. 25.

**4-1  $P$ = 100 bar**

The MC simulations were performed at $P$ = 100 bar and $T$ = 220, 240, 260, 280, and 300 K, as shown in Fig. 1. At this pressure, the melting temperature is $T_m$ = 218.6 K. The state {$P$ = 100 bar, $T$ = 300 K} is close to the critical point. The Frenkel anomaly observed in the present study was {$P$ = 100 bar, $T$ = 260 K}, which was in good agreement with the results reported in Ref. 10.

**4-1-1  Molar volume**

Computed molar volumes are an important indicator in order to know to what degree the empirical molecular potential model and the simulation program used are reliable. In Fig. 3, the computed molar volumes (dark points) are shown together with the corresponding NIST values (red points) [25]. In what follows, the dark and red lines in the figure are shown only for eye-guide. The computed molar volume was 2.2% greater at $T$ = 220 K and 0.9% greater at $T$ = 300 K than the corresponding NIST values. These deviations were considered to be small enough to study liquid $CO_2$. In Fig. 3, the vertical blue line indicates $T_c$, and hence the region on the left-hand side of the blue line is the normal fluid region, while the opposite side is the SC liquid region. The data points in the SC liquid region at $T$ = 310 and 320 K were obtained only by preliminary MC simulations. At these two temperatures, the computed molar volumes were significantly smaller than the corresponding NIST values, indicating that the computational protocol of MC simulations used in the present study failed to reproduce the SC liquid state in the region {$P$ = 100 bar, $T$ > 300 K} partly because the fluctuations of liquid $CO_2$ parameters, computed by a *NPT*-MC simulation, are in general greater than those computed by a *NVT*-MC simulation [21].

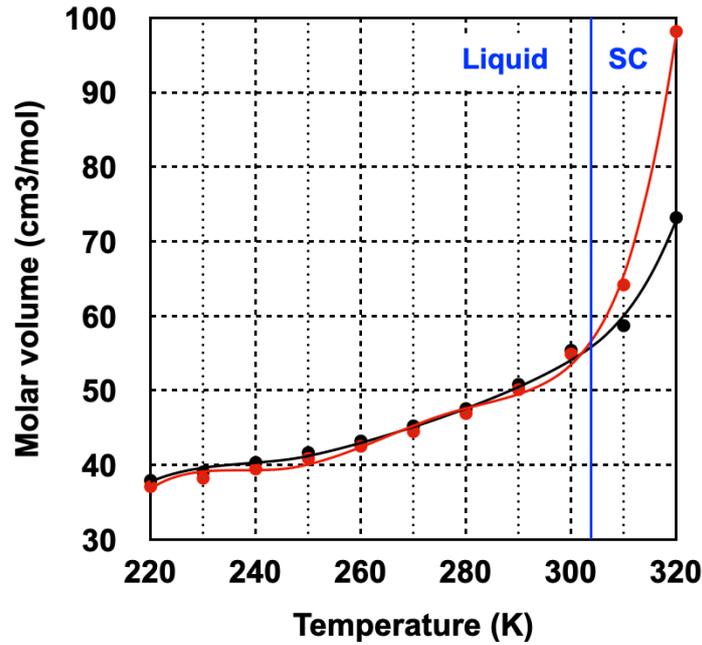

Fig. 3. Computed molar volume (dark dots) in comparison with the NIST data (red dots)[25] at $P$ = 100 bar. The vertical blue line shows the location of $T_C$, hence the region on the left-hand side of the blue line is the normal liquid while the region on the right-hand side is the SC liquid.

**4-1-2 Radial distribution function between molecular centers**

Figure 4 shows the RDFs between molecular centers at $T$ = 230 K (dark line), 260 K (blue line), and 290 K (red line). Note that the notations, $g_{ll'+m/-m}$, used in the previous study [17] are simply denoted as $g(ll'm)$ in the present article without any loss of information, and $g(000)$ is equivalent to the conventional notation $g(r)$. The notations $g(000)$ and $g(r)$ will be used interchangeably hereafter. The RDF was computed up to $r$ = 20 Å with a step size of 0.1 Å. It should be noted that the Kihara potential works up to 10 Å, while both the van der Waals potential and the EQQ potential works up to 15 Å. Therefore, no intermolecular potential acts beyond 15 Å; therefore, $g(r)$ in the range from 15 to 20 Å corresponds to the background value of $g(r)$ = 1.0. The small fluctuation beyond 15 Å was attributed to numerical noise. In $g(000)$ of Fig. 4, the first peak height of the RDF was 1.90, and the first minimum position was $r$ = 5.85 Å at $T$ = 220 K. The peak height lowered smoothly with $T$, reaching 1.66 at $T$ = 300 K.

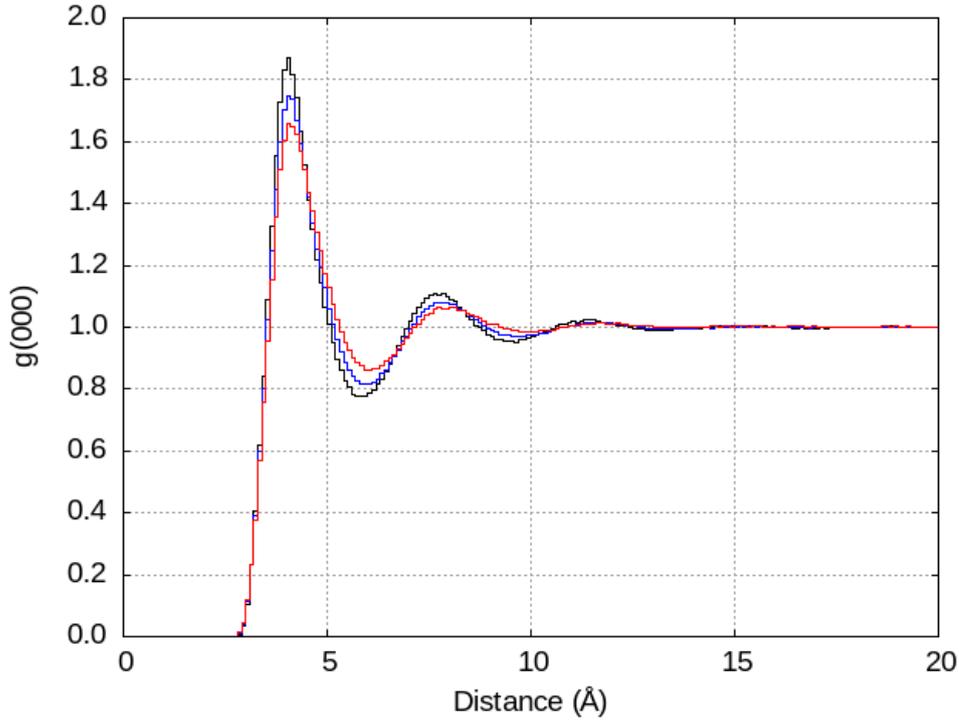

Fig. 4. Radial distribution functions between molecular centers, $g(000)$ at {$P$ = 100 bar, $T$ = 230 K (dark line), 260 K (blue line), and 290 K (red line)}.

### 4-1-3 Co-ordination numbers

The co-ordination number, $CN$, is defined by:

$$CN = 4\pi\rho_N \int_0^{r(first\ minimum)} r^2\, g(000)\, dr \qquad (3)$$

Here, $\rho_N$ denotes the number density, $N/V$, and $g(000)$ is equivalent to $g(r)$. In Eq. 3, the integration is undertaken from $r$ = 0 Å to $r$ = first minimum between the first and second peaks of $g(r)$. Therefore, the $CN$ represents the average number of molecules located within the closest molecular *shell* surrounding a central molecule.

Figure 5 shows the $CNs$ as a function of $T$, as computed in the present study. The $CN$ curve begins to drop around $T$ = 260 K, and from this results, it was concluded that the Frenkel anomaly at $P$ = 100 bar occurred at $T$ = 260 K, which was in good agreement with the result reported in Ref. 10. The $CN$ values below $T$ = 260 K were slightly larger than that of the cubic $CO_2$ phase I structure. ($CN$ = 12). This is presumably because $CO_2$ molecules tend to align unidirectionally as the $CO_2$ molecule has a rod shape (see Sec. 4) [27], and hence the $CO_2$ molecules were more densely packed in fluid than in solid phase I.

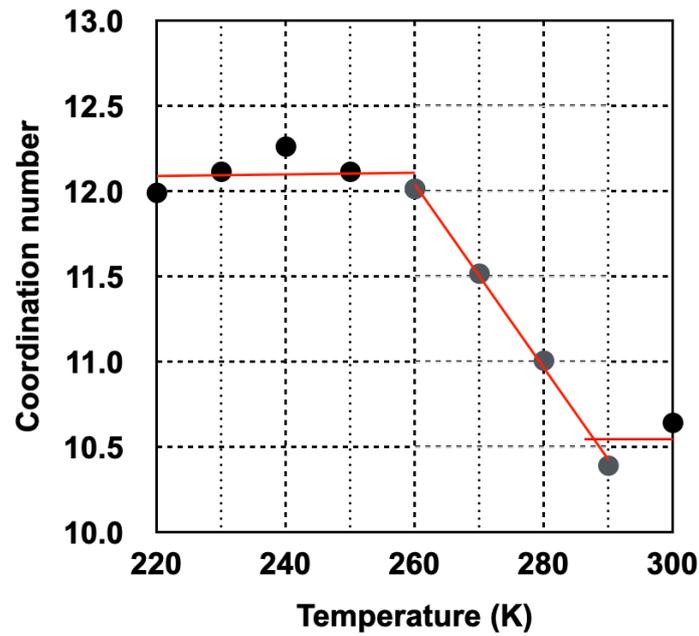

Fig. 5. Co-ordination numbers at $P$ = 100 bar between $T$ = 220 and 300 K. The red lines cross at about $T$ = 260 and 288 K.

### 4-1-4 First minimum positions in $g(r)$

The first minimum position in $g(r)$ at each temperature is shown in Fig. 6. It appears that the first minimum position begins to increase rapidly at $T$ = 260 K, corresponding to the temperature at which the Frenkel anomaly occurred. Note that the mesh size in $g(r)$ was 0.1 Å so that the error bar was at least greater than 0.1 Å. Furthermore, the first minimum itself was very shallow. These two factors made it difficult to more precisely determine the first minimum position in $g(r)$.

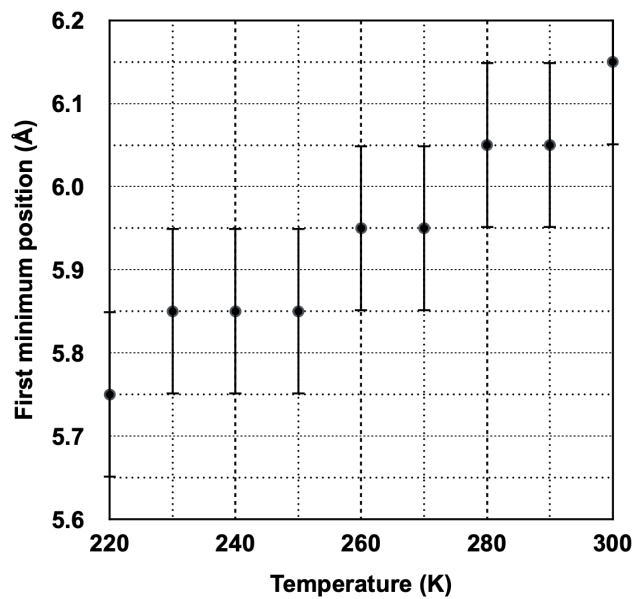

Fig. 6. First minimum positions of RDF at different temperatures at $P$ = 100 bar. The mesh size is 0.1 Å, and the error bars, ±0.1 Å, are very approximate. Actually, they are larger at higher temperature.

**4-1-5 *CNs* for *g*(220)**

In Ref. 17, the RDF at {$P$ = 50 bar, $T$ = 250 K: normal liquid} was computed as well as the expansion coefficients $g(ll'm)$ of the RDF by spherical harmonics as a function of intermolecular distance up to 20 Å for all possible combinations of angular indices $l$, $l'$, and $|m|$ ≤ 8. It is of interest to see how $g(ll'm)$ changes with $T$. Among many $g(ll'm)$, $g(220)$ was examined here. As an example, $g(220)$ at $T$ = 260 K is shown in Fig. 7. Here, it should be noted that:

$$g(220) \approx < Y_{2,0}(\omega_1) Y_{2,0}(\omega_2) > \tag{4}$$

where the angles $\omega_1$ and $\omega_2$ refer to the axis passing through the molecular centers of molecules 1 and 2, and $Y_{2,0}(\omega)$ is a spherical harmonics with $l$ = 2 and $m$ = 0:

$$Y_{2,0}(\theta, \varphi) = \frac{1}{4}\sqrt{\frac{5}{\pi}}(3\cos^2\theta - 1) \tag{5}$$

Therefore, $g(220)$ is supposed to express the orientational correlation between the central and the nearest neighbor molecules. A band shape of $g(220)$ at $T$ = 260 K is shown below in Fig. 7. The band shapes for other temperatures were similar.

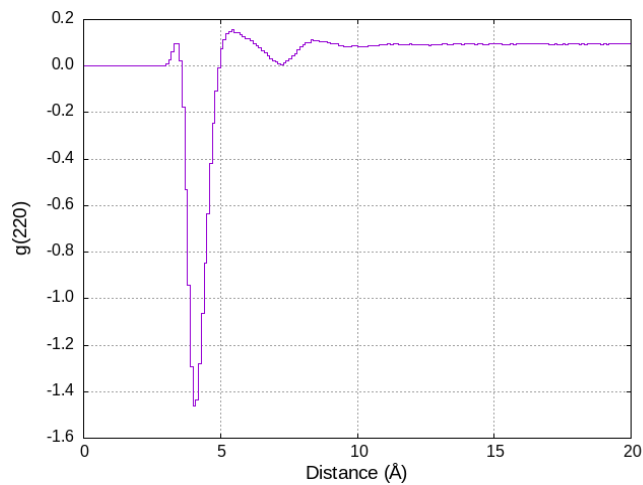

Fig. 7. Band shape of $g(220)$ at $T$ = 260 K.

The *CNs* associated with the first negative band in $g(220)$ were computed in the same manner as the *CNs* for $g(000)$ was computed in Sec. 3-4, and the result is shown in Fig. 8. The two red lines appears to cross at $T$ = 260 K where the Frenkel anomaly took place.

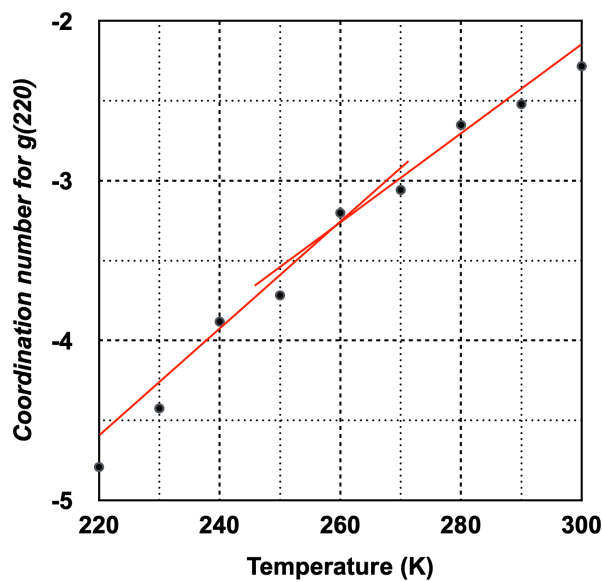

Fig. 8. Co-ordination numbers for the first negative band in $g(220)$ against $T$.

Figure 8 indicates that the orientational order, represented by the *CN* of $g(220)$, becomes increasingly random as $T$ increases, and the slope of the *CN* decreases slightly in crossing the Frenkel line.

**4-2 *P* = 1000 bar**

The MC simulations were performed at $P$ = 1000 bar and temperatures spaced every 20 K between 240 and 420 K, as indicated by red dots and a circle in Fig. 1. These {*P*, *T*} states are in the SC liquid region, and at this pressure, the melting temperature $T_m$ = 235 K. The Frenkel anomaly was found at {*P* = 1000 bar, *T* = 370 K} in the present study. This {*P*, *T*} state differs significantly from that reported in Ref. 10, where the Frenkel anomaly is located at {*P* = 1000 bar, *T* = 320 K}.

**4-2-1  Molar volume**

In Fig. 9, the computed molar volumes (dark points) are shown together with the corresponding NIST values (red points) [25]. The computed molar volumes were 1.5 - 1.8 % higher than the NIST values, which is considered to be sufficiently small for the subsequent *CN* computations.

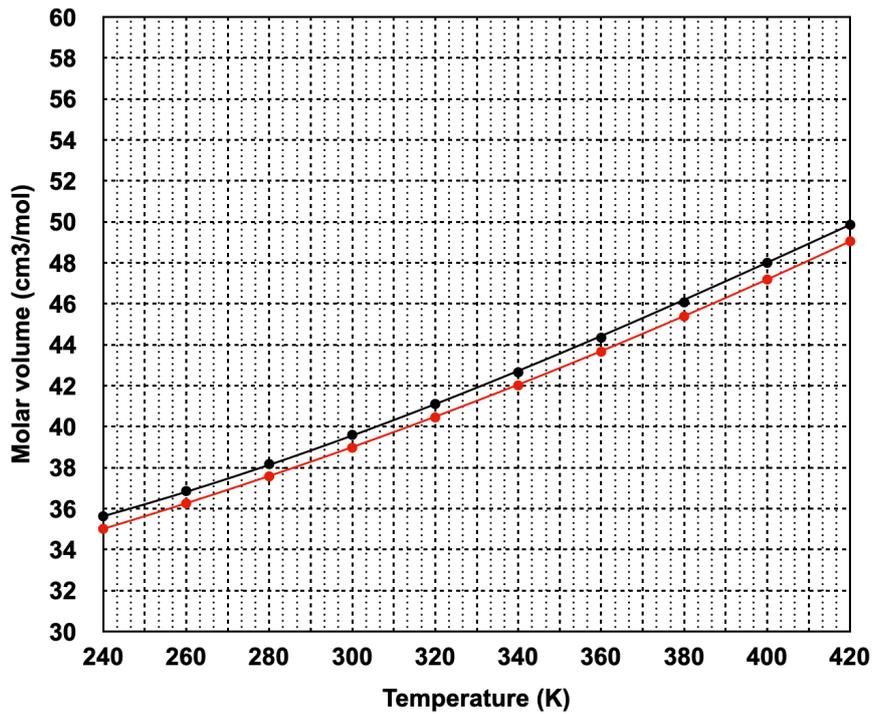

Fig. 9. Computed molar volume (dark dots) in comparison with the NIST data (red dots)[25] at $P$ = 1000 bar. The liquid $CO_2$ is in the SC liquid region.

### 4-2-2 Radial distribution function between molecular centers

Figure 10 shows the RDF at $T$ = 240 K (dark line), 360 K (blue line), and 420 K (red line) between molecular centers. The first peak height of the RDF was 1.92, and the first minimum position was $r$ = 5.65 Å at $T$ = 240 K. The peak height lowered smoothly with $T$, reaching 1.51 at $T$ = 420 K.

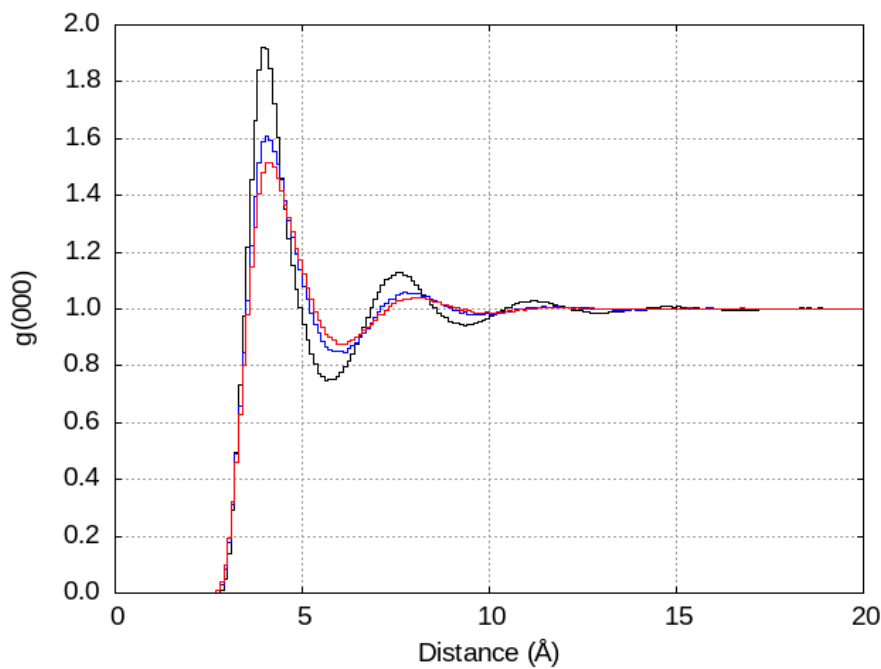

Fig. 10. Radial distribution functions between molecular centers, $g(000)$, at {$P$ = 1000 bar, $T$ = 240 K (dark line), 360 K (blue line), and 420 K (red line)}.

**4-2-3 Co-ordination numbers**

Figure 11 shows the *CNs* as a function of $T$, computed in the present study. It is observed that the *CN* curve begins to drop at approximately $T$ = 370 K, and becomes nearly constant above $T$ = 410 K. Therefore, it was concluded that the Frenkel anomaly at $P$ = 1000 bar occurred at $T$ = 370 K. This temperature was 50 K higher than that reported in Ref. 10, where the Frenkel anomaly is located at 320 K. Namely, at $P$ = 1000 bar, the computed Frenkel anomaly was located as far as 135 K away from the melting temperature $T_m$ = 235 K. This means that liquid $CO_2$ behaves as a rigid liquid over a large $T$ region of 135 K.

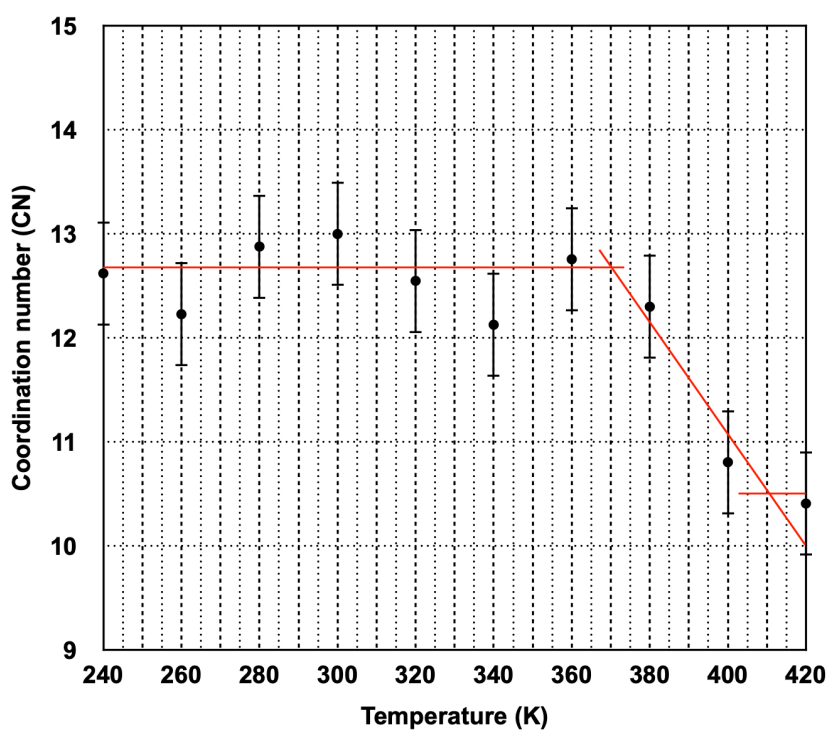

Fig. 11. Co-ordination numbers at $P$ = 1000 bar between $T$ =240 and 420 K. The two red lines crossing at about $T$ = 370 K indicates the location of the Frenkel anomaly.

**4-2-4 First minimum positions in RDF**

The first minimum position in the RDF at each temperature is shown in Fig. 12. It appears that the first minimum position initially increases with temperature but becomes nearly constant at $T \geq 350$ K. This behavior differs markedly from that at $P$ = 100 bar (Fig. 6). Note again that like in the case of $P$ = 100 bar, the mesh size of $g(r)$ in the present computations was 0.1 Å so that the error bar was at least greater than 0.1 Å. Moreover, the first minimum itself was very shallow. These two factors made it difficult to more precisely determine the first minimum position in $g(r)$.

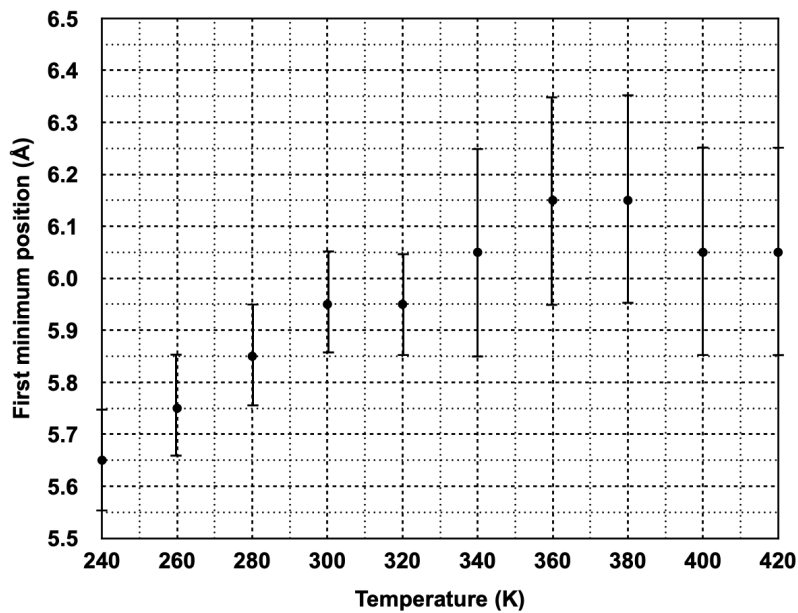

Fig. 12. First minimum position of RDF at different temperatures at $P$ = 1000 bar. The length of the error bars are very approximate.

**4-2-5 *CNs* for *g*(220)**

Figure 13 shows the band shape of $g(220)$ at $T$ = 360 K, and the co-ordination numbers for $g(220)$ were computed in the same manner as in Sec. 3-2-5.

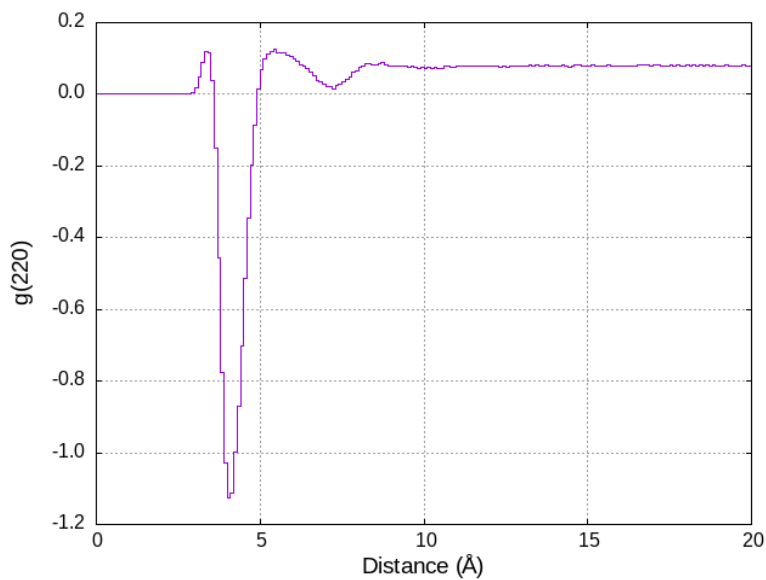

Fig. 13. Band shape of $g(220)$ at $T$ = 360 K.

The $T$-dependence of the $CN$ for $g(220)$ was computed and the result is shown in Fig. 14. Unlike the case of $P$ = 100 bar, no change was visible at $T$ = 370 K where the Frenkel anomaly took place.

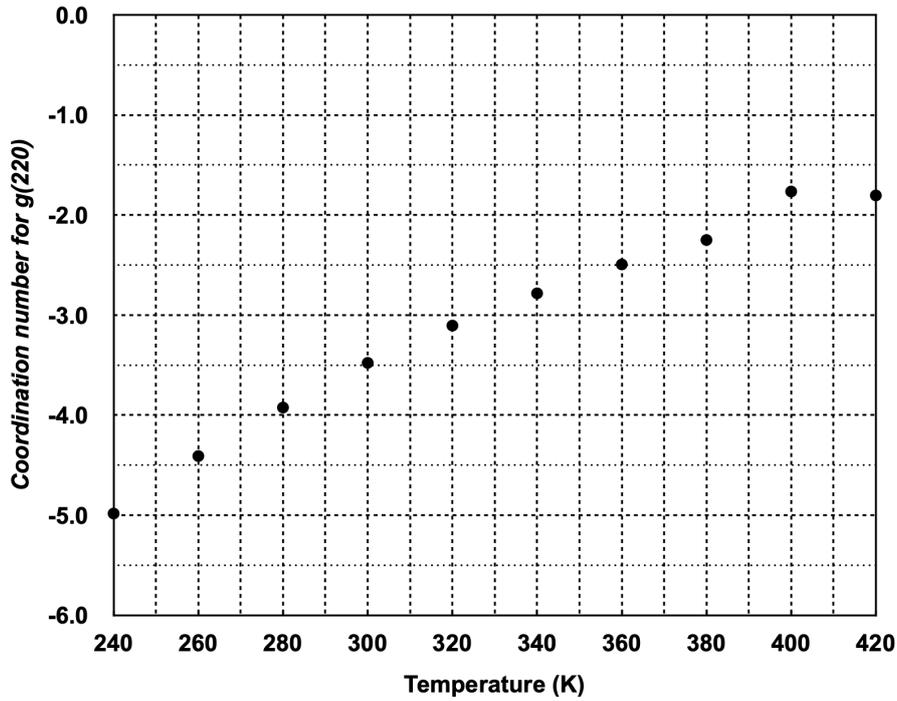

Fig. 14. Co-ordination numbers for $g(220)$ at $P$ = 1000 bar between $T$ = 240 and 420 K.

## 5. Molecular orientations

In the previous study [17], the molecular directions were projected onto a hemisphere, and it was found that $CO_2$ molecules tend to align along the z-direction (essentially, along a single direction). This tendency was re-examined in the present study at {$P$ = 100 bar, $T$ = 260 K} and {$P$ = 1000 bar, $T$ = 360 K}, as shown in Fig. 15. The results similarly indicate that $CO_2$ molecules tend to align along one direction. Idrissi *et al* [27] also reported that the closer neighbors have a higher probability of adopting parallel orientations. The present results therefore suggest that liquid $CO_2$ consists of domains in which $CO_2$ molecules tend to be oriented in one direction. The domain size should be larger than 50 Å, which is an approximate edge length of the cubic box used in the present MC simulations. As seen in Figs. 8 and 14, there was no significant change in the *CN*s of $g(220)$ at the Frenkel anomaly, even though a substantial transition from a rigid liquid to a normal fluid occurs. Further investigation of the orientational behavior along the Frenkel line is therefore required.

(a)

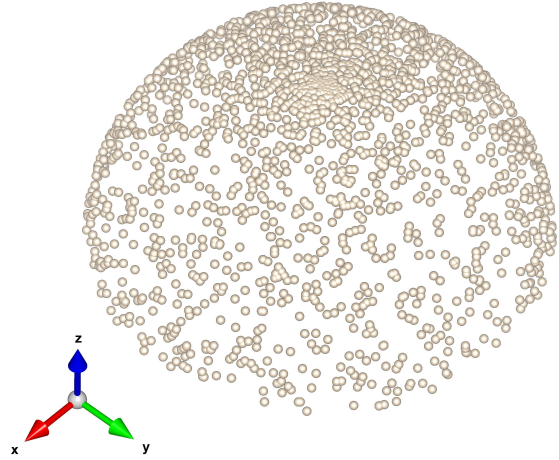

(b)

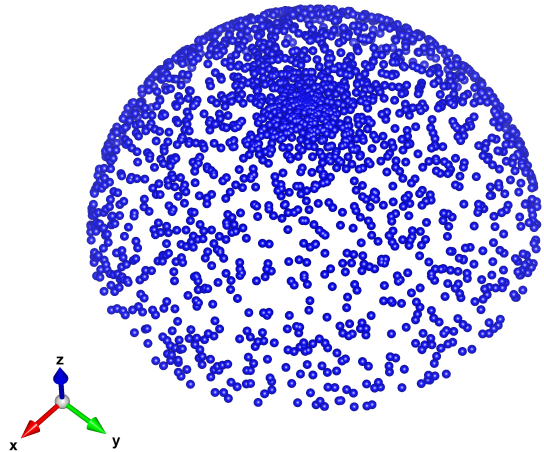

Fig. 15.  Projection of molecular directions on a hemi-sphere: (a) {$P$ = 100 bar, $T$ = 260 K}, (b) {$P$ = 1000 bar, $T$ = 360 K} [28].

## 6. Conclusion

The *CNs* were computed by MC simulation using the Kihara potential over the ranges {$P$ = 100 bar, $T$ = 220 - 300 K: the normal fluid region} and {$P$ = 1000 bar, $T$ = 240 - 420 K: the SC liquid region}. An anomalous behavior was identified at {$P$ = 100 bar, $T$ = 260 K}, which is most likely a Frenkel anomaly, as this position was in good agreement with the result reported in Ref. 10. A similar anomaly in *CN* was also observed at {$P$ = 1000 bar, $T$ = 370 K} although it differed significantly from that reported in Ref. 10. In contrast, the *CNs* for *g*(220) exhibited a smooth increase with $T$ in the two cases, suggesting that in the rigid liquid, molecular orientations become as random as an ordinary liquid with $T$. In conclusion, the present study confirmed that the MC simulation is a very useful tool for identifying the Frenkel anomaly in liquid $CO_2$. The Frenkel line/anomaly is a peculiar phenomenon whose physical cause has not been understood, and even its precise location remains uncertain despite numerous studies. Addressing and resolving these issues will be a challenging but important task for future research.

**Acknowledgment**

The author thanks Keiko Nishikawa for providing a number of useful information on supercritical fluids, particularly on the Nishikawa line and fluctuations.